\documentclass[reprint,aps,prl,showpacs,superscriptaddress]{revtex4-2}

\usepackage{CJK}
\usepackage[tbtags]{amsmath}
\usepackage{amssymb}
\usepackage{bmpsize}
\usepackage{mathtools}
\usepackage{caption}
\usepackage{amsfonts}    
\usepackage{graphicx}   
\usepackage{verbatim}   
\usepackage{color}      
\usepackage{subfigure}  
\usepackage{hyperref}   
\usepackage{natbib}
\usepackage{booktabs}
\usepackage{threeparttable}
\usepackage{dcolumn}
\usepackage{multirow}
\usepackage{fancyhdr}

\makeatletter

\newcommand{\Rmnum}[1]{\expandafter\@slowromancap\romannumeral #1@}
\makeatother

\begin{document}

\date{\today}

	\begin{CJK}{UTF8}{gbsn}
	\title{Experimental realization of a three-photon asymmetric maximally entangled state and its application to quantum teleportation}

\author{Linxiang Zhou}
\altaffiliation{These authors contributed equally}
\affiliation{State Key Laboratory of Optoelectronic Materials and Technologies and School of Physics, Sun Yat-sen University, Guangzhou 510000, People's Republic of China}

\author{Qiao Xu}
\altaffiliation{These authors contributed equally}
\affiliation{State Key Laboratory of Optoelectronic Materials and Technologies and School of Physics, Sun Yat-sen University, Guangzhou 510000, People's Republic of China}

\author{Tianfeng Feng}
\altaffiliation{These authors contributed equally}
\affiliation{State Key Laboratory of Optoelectronic Materials and Technologies and School of Physics, Sun Yat-sen University, Guangzhou 510000, People's Republic of China}

\author{Xiaoqi Zhou}\email{zhouxq8@mail.sysu.edu.cn}
\affiliation{State Key Laboratory of Optoelectronic Materials and Technologies and School of Physics, Sun Yat-sen University, Guangzhou 510000, People's Republic of China}
\affiliation{Hefei National Laboratory, University of Science and Technology of China, Hefei 230088, China}

	\begin{abstract}
	Quantum entanglement is a fundamental resource for quantum information processing and is widely used in quantum communication, quantum computation and quantum metrology. Early research on quantum entanglement mainly focus on qubit states, but in recent years, more and more research has begun to focus on high-dimensional entangled states. Compared with qubit entangled states, higher-dimensional entangled states have a larger information capacity and the potential to realize more complex quantum applications. In this Letter, we have experimentally prepared a special high-dimensional entangled state, the so-called three-photon asymmetric maximally entangled state, which consists of two two-dimensional photons and one four-dimensional photon. Using this asymmetric maximally entangled state as a resource, we have also implemented a proof-of-principle quantum teleportation experiment, realizing the transfer of quantum information from two qubits to a single ququart. The fidelities of the quantum teleportation range from 0.79 to 0.86, which are well above both the optimal single-copy ququart state-estimation limit of $2/5$ and maximal qutrit-ququart overlap of $3/4$, thus confirming a genuine and nonclassical four-dimensional teleportation. The asymmetric entangled state realized here has the potential to be used as a quantum interface in future quantum networks, allowing quantum information transfer between quantum objects of different dimensions via the quantum teleportation protocol demonstrated in this work.
\end{abstract}

%
%

%

\maketitle	

\emph{Introduction.}---Quantum entanglement \cite{horodecki2009quantum,friis2019entanglement} is one of the most fundamental and distinctive features of quantum mechanics, with
many applications in the field of quantum information science, such as quantum teleportation \cite{pirandola2015advances,bennett1993teleporting,bouwmeester1997experimental,wang2015quantum}, quantum metrology \cite{giovannetti2011advances,giovannetti2004quantum} and quantum communication \cite{hillery1999quantum,cerf2002security,duan2001long}.
The most basic and fundamental quantum entangled state is the two-particle, two-dimensional state $\frac{1}{\sqrt{2}}\left( |00\rangle +|11\rangle\right)$, which can be used to show the contradiction between quantum mechanics and local realism \cite{bell1964einstein,Hardy1992,Hardy1993} and as a physical resource for realizing quantum teleportation of a two-dimensional quantum state.
More advanced quantum applications require more complex quantum entangled states. To increase the complexity of quantum entangled states, researchers have worked in two directions: increasing the number of particles and increasing the dimensionality of the particles. Take optical system as an example, by increasing the number of photons, a variety of multi-photon two-dimensional entangled states have been prepared, such as 10-photon \cite{wang2016experimental,chen2017observation} and 12-photon entangled states \cite{zhong201812}. In terms of increasing the dimensionality of photons, a variety of two-photon high-dimensional entangled states have also been generated \cite{erhard2020advances,wang2018multidimensional,chang2021648,hu2020efficient}, such as the $100\times100$ orbital angular momentum (OAM) entangled state \cite{krenn2014generation}. 
By using such high-dimensional entangled states as the physical resources, quantum teleportation of a high-dimensional quantum state has also been demonstrated \cite{hu2020experimental,luo2019quantum}.

To further increase the complexity of quantum entangled states, attempts have been made to increase the number and dimensionality of particles simultaneously, i.e., to prepare multi-particle high-dimensional quantum entangled states. So far, several works on three-particle high-dimensional entangled states have been reported, including the preparation of the (3, 3, 2) state $
\frac{1}{\sqrt{3}}\left( |000\rangle +|111\rangle +|221\rangle \right)
$ \cite{malik2016multi}, the (3, 3, 3) state $
\frac{1}{\sqrt{3}}\left( |000\rangle +|111\rangle +|222\rangle \right) 
$  \cite{erhard2018experimental} and the (4, 4, 2) state $
\frac{1}{2}\left( |000\rangle +|111\rangle +|220\rangle +|331\rangle \right)
$ \cite{hu2020experimental2}, which hold promise for applications in some special quantum communication networks.

In this Letter, we report the preparation of a three-photon asymmetric maximally entangled state, which is the (2, 2, 4) state $\frac{1}{2}\left( |000\rangle+|011\rangle+|102\rangle+|113\rangle \right)$.
Such asymmetric maximally entangled state can be used to realize a quantum teleportation operation between particles of different dimensions, such as teleporting an unknown quantum state with two qubits of quantum information from two two-dimensional particles to one four-dimensional particle and vice versa. 
In addition to preparing this asymmetric maximally entangled state, we have also performed a proof-of-principle experiment using it as the physical resource to demonstrate the quantum teleportation from two qubits to a single ququart. Our methods have the potential to be applied to future quantum networks for quantum information transfer between quantum objects of different dimensions.

\begin{figure}
	\centering
	\includegraphics[scale=0.23
	]{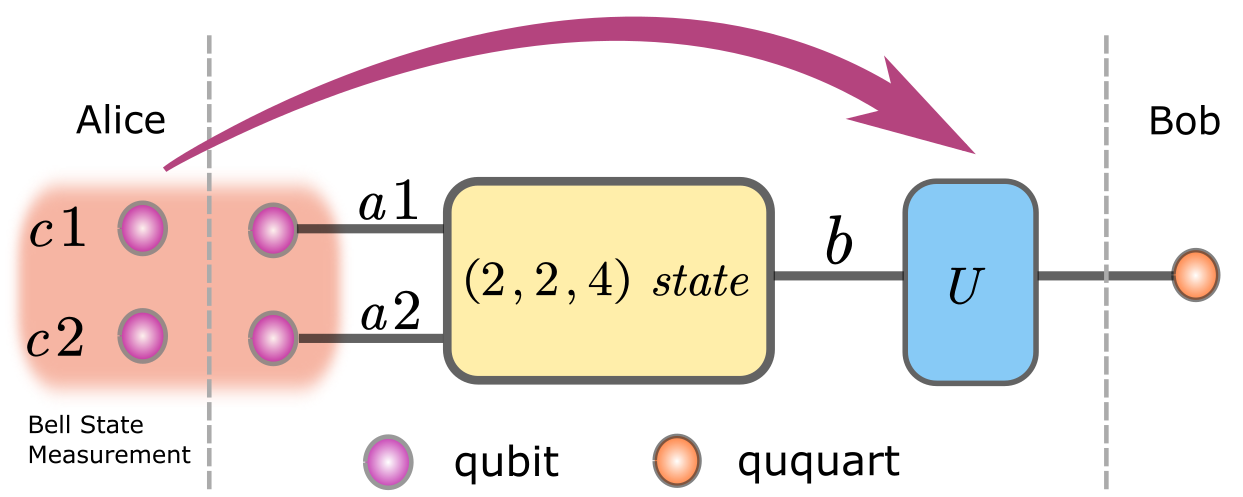}
	\captionsetup{justification=raggedright}
	\caption{Quantum teleportation protocols based on the (2, 2, 4) state. By sending qubits $a1$ and $a2$ to Alice and ququart $b$ to Bob, a quantum teleportation can be realized between Alice and Bob. If Alice has two qubits $c1$ and $c2$, she can make Bell state measurements on qubits $a1$ and $c1$ and on qubits $a2$ and $c2$ respectively, and then send the measurement results to Bob. Based on the measurement results, Bob then performs the corresponding unitary transformation on $b$, thus realizing the quantum teleportation from two qubits $c1$ and $c2$ to ququart $b$.} 
		%
\end{figure}

\emph{The (2, 2, 4) state and its applications on quantum teleportation.}---Before introducing the asymmetric maximally-entangled state, we first review the standard maximally-entangled state. 
For example, the (4, 4) state is a standard maximally-entangled state consisting of two four-dimensional particles $a$ and $b$
\begin{equation}
|{4,4}\rangle =\frac{1}{2}\left( |00\rangle_{a,b} +|11\rangle_{a,b}+|22\rangle_{a,b}+|33\rangle_{a,b} \right), 
\end{equation}
which can be used as a physical resource for quantum teleportation of a four-dimensional quantum state \cite{bennett1993teleporting}.

The (2, 2, 4) state is a related quantum state 
\begin{equation}
		\begin{aligned}
		|2,2,4\rangle=
		&\frac{1}{2}(|000\rangle _{a1,a2,b}+|011\rangle _{a1,a2,b}
		\\
		&+|102\rangle _{a1,a2,b}+|113\rangle _{a1,a2,b}),
		\end{aligned}
	\end{equation}
where $a1$ and $a2$ are two-dimensional particles, and $b$ is a four-dimensional particle.
Comparing Eq. (2) and Eq. (1), it can be seen that there is a direct correspondence between the (2, 2, 4) state and the (4, 4) state.
For the (2, 2, 4) state, if particle $a1$ and $a2$ are regarded as system A and particle $b$ is regarded as system B, both A and B are four-dimensional systems, and their composite system is in a four-dimensional maximally-entangled state. Since the number of particles in systems A and B are different, we call the (2, 2, 4) state an asymmetric maximally-entangled state. 

%



Similar to the symmetric (4, 4) state, the asymmetric (2, 2, 4) state can also be used as a physical resource for quantum teleportation. By sending particle $a1$ and $a2$ to Alice and particle $b$ to Bob, the (2, 2, 4) state can enable quantum teleportation of a four-dimensional quantum state from Alice to Bob. 

As shown in Fig. 1, suppose Alice has two-dimensional particles $c1$ and $c2$ in an unknown quantum state
\begin{equation}
	\alpha |00\rangle _{c1,c2}+\beta |01\rangle _{c1,c2}+\gamma |10\rangle _{c1,c2}+\delta |11\rangle _{c1,c2}, 
\end{equation}
which can be either a separable or entangled state. 
The joint quantum state of the five particles $c1$, $c2$, $a1$, $a2$ and $b$ can thus be written as
\begin{equation}
	\begin{aligned}
   & (\alpha |00\rangle _{c1,c2}+\beta |01\rangle _{c1,c2}+\gamma |10\rangle _{c1,c2}+\delta |11\rangle _{c1,c2})\\
   \otimes &\frac{1}{2}(|000\rangle _{a1,a2,b}+|011\rangle _{a1,a2,b}+|102\rangle _{a1,a2,b}+|113\rangle _{a1,a2,b})\\
      =&|\Phi ^+\rangle _{c1,a1}|\Phi ^+\rangle _{c2,a2}\otimes\left(\alpha |0\rangle_b +\beta |1\rangle_b +\gamma |2\rangle_b +\delta |3\rangle_b \right) 
      \\
      +&|\Phi ^+\rangle _{c1,a1}|\Phi ^-\rangle _{c2,a2}\otimes\left(\alpha |0\rangle_b -\beta |1\rangle_b +\gamma |2\rangle_b -\delta |3\rangle_b \right)
      \\
      &\cdots \cdots
      \\
      +&|\Psi ^-\rangle _{c1,a1}|\Psi ^+\rangle _{c2,a2}\otimes\left(-\delta  |0\rangle_b -\gamma |1\rangle_b + \beta|2\rangle_b +\alpha|3\rangle_b \right) 
      \\
      +&|\Psi ^-\rangle _{c1,a1}|\Psi ^-\rangle _{c2,a2}\otimes\left(\delta  |0\rangle_b -\gamma |1\rangle_b - \beta|2\rangle_b +\alpha|3\rangle_b \right) ,
    \end{aligned}
\end{equation}
where $|\Phi ^\pm\rangle=\frac{1}{\sqrt{2}}\left( |0\rangle |0\rangle \pm |1\rangle |1\rangle \right)$ and $|\Psi ^{\pm}\rangle =\frac{1}{\sqrt{2}}\left( |0\rangle |1\rangle \pm |1\rangle |0\rangle \right)$.
Alice performs Bell state measurements on the two particles $c1$ and $a1$ and the two particles $c2$ and $a2$, respectively, and sends the measurement results to Bob via classical channel. Bob then performs the corresponding local unitary operation on particle $b$ based on the measurement results, thus obtaining
\begin{equation}
	\alpha |0\rangle_b +\beta |1\rangle_b +\gamma |2\rangle_b +\delta |3\rangle_b.
\end{equation}
Comparing Eq. (5) with Eq. (3), it can be seen that the two quantum states have exactly the same form except for the difference in the state basis, which indicates that the unknown quantum state with two qubits of quantum information has been successfully teleported from the two-dimensional particles $c1$ and $c2$ to the four-dimensional particle $b$.  

Similarly, Bob can also use this (2, 2, 4) state as a resource to teleport a four-dimensional quantum state from a four-dimensional particle to two two-dimensional particles (See Supplementary Materials \Rmnum{1}).

	\begin{figure*}
	\centering
	\includegraphics[scale=0.3]{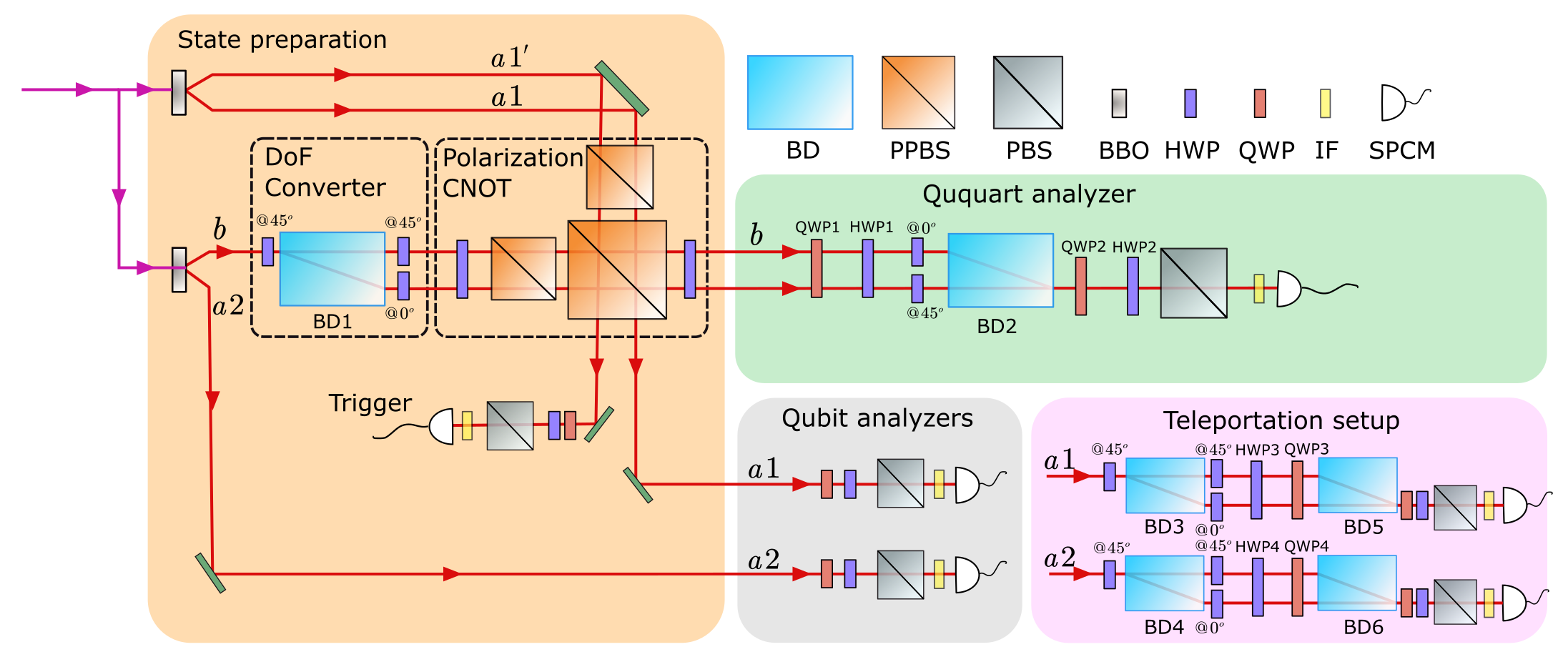}
	\captionsetup{justification=raggedright}
	\caption{Experimental setup for the preparation of the (2, 2, 4) state. An ultraviolet pulsed laser is focused on two beta-barium borate (BBO) crystals to produce two photon pairs $a1'$-$a1$ and $b$-$a2$. 
		After passing through the degree-of-freedom (DoF) converter, photon $b$ is split into the upper and lower paths. The upper (lower) path of photon $b$ is then interacted with photon $a1'$ ($a1$) through a polarization controlled-NOT (CNOT) gate.
		Photon $a1'$ is then triggered, and the remaining three-photon state consisting of photons $a1$, $a2$ and $b$ becomes the desired (2, 2, 4) state. Here the experimentally obtained fourfold coincidence count rate is 0.183 Hz.
		In order to analyze this (2, 2, 4) state, photon $b$ is fed to the ququart analyzer (green region) and photons $a1$ and $a2$ are fed to the qubit analyzers (gray region).
		In the case of performing quantum teleportation experiment based on the (2, 2, 4) state, photons $a1$ and $a2$ are directly fed into the teleportation setup (pink region) instead of entering the qubit analyzers. 		
		BBO, beta-barium borate; BD, beam displacer; PPBS, partial polarization beamsplitter; PBS, polarization beamsplitter; HWP, half waveplate; QWP, quarter waveplate; IF, interference filter; SPCM, single-photon counting modules.}
\end{figure*}

\begin{figure*}[htb]
	\centering
	\includegraphics[scale=0.16]{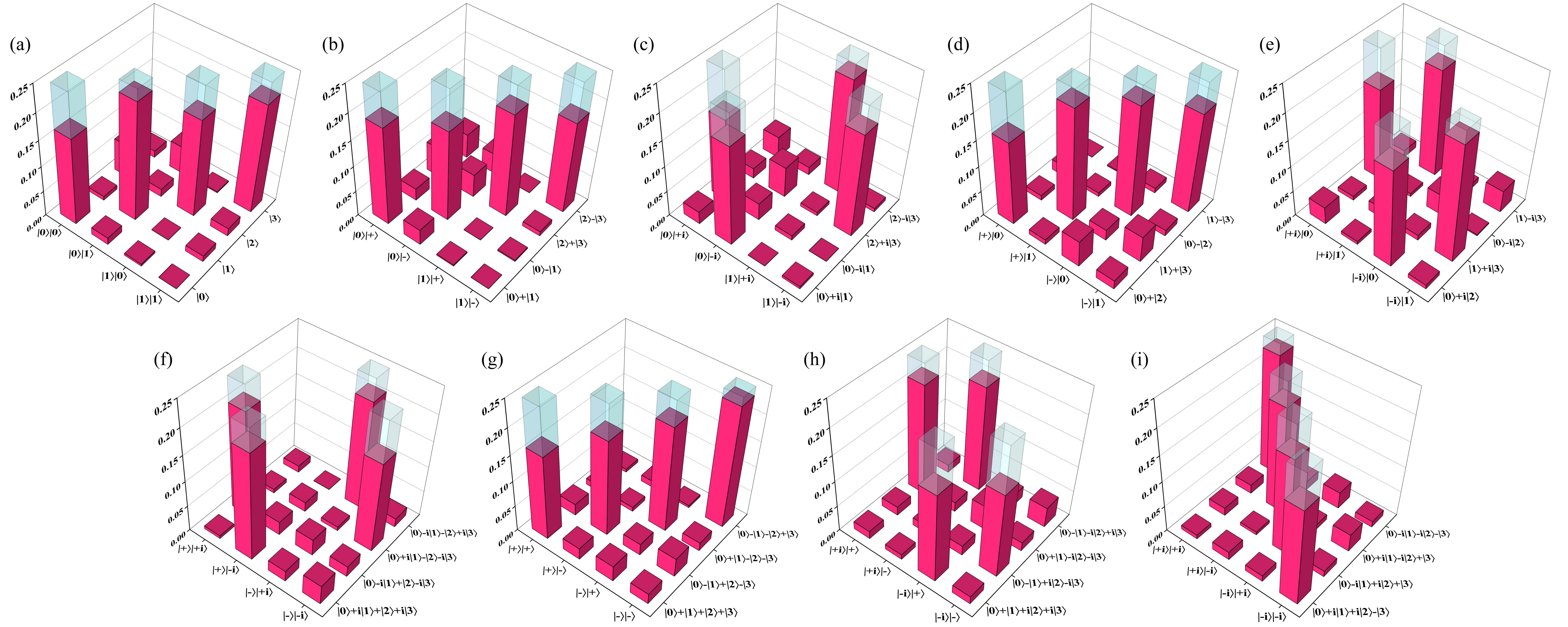}
	\captionsetup{justification=raggedright}
	\caption{The measurement results of the prepared three-photon state at nine different bases. The fidelity of the prepared state with respect to the ideal (2, 2, 4) state can be calculated from these measurement results as $0.72\pm0.02$.
		%
		%
		%
	}
\end{figure*}

\emph{Experimental results}---
Here we experimentally prepared a three-photon entangled state
\begin{equation}
		\begin{aligned}
			&\frac{1}{2}(|H\rangle _{a1} |H\rangle _{a2}|Hu\rangle_{b}+ |H\rangle _{a1}|V\rangle _{a2}|Hl\rangle_{b}\\
			&+|V\rangle _{a1} |H\rangle _{a2}|Vu\rangle_{b}+ |V\rangle _{a1}|V\rangle _{a2}|Vl\rangle_{b}).
		\end{aligned}
	\end{equation}	
where $H$ ($V$) denotes the horizontal (vertical) polarization, $Hu$ $(Hl)$ denotes horizontal polarization in the upper (lower) path, and $Vu$ $(Vl)$ denotes vertical polarization in the upper (lower) path. 
Photons a1 and a2 are two-dimensional as they use only the polarization degree of freedom (DoF), and photon b is four-dimensional as it uses both polarization and path DoFs.
The experimental setup is shown in Fig. 2, and the experimental details are provided in the Supplementary Materials \Rmnum{2}. 

By defining $|H\rangle\equiv|0\rangle$ and $|V\rangle\equiv|1\rangle$ for photons $a1$ and $a2$, and $|Hu\rangle\equiv|0\rangle$, $|Hl\rangle\equiv|1\rangle$, $|Vu\rangle\equiv|2\rangle$ and $|Vl\rangle\equiv|3\rangle$ for photon $b$, the prepared three-photon state can be rewritten as
	\[
	\frac{1}{2}(|000\rangle _{a1,a2,b}+|011\rangle _{a1,a2,b}+|102\rangle _{a1,a2,b}+|113\rangle _{a1,a2,b}),
	\]
	which is exactly our desired (2, 2, 4) state.

To evaluate the fidelity of the prepared (2, 2, 4) state, we let photon $b$ enter a device (green region) for measuring ququart states (See Supplementary Materials \Rmnum{3}), and photons $a1$ and $a2$ enter a device (gray region) for measuring qubit states.
We then performed measurements on the prepared three-photon state under nine different measurement bases (See Supplementary Materials \Rmnum{4}), and the results are shown in Fig. 3. Based on these measurement results, the fidelity of the prepared three-photon state with respect to the ideal (2, 2, 4) state can be calculated to be $0.72\pm0.02$.

	
 \begin{figure*}
	\centering
	\includegraphics[scale=0.26]{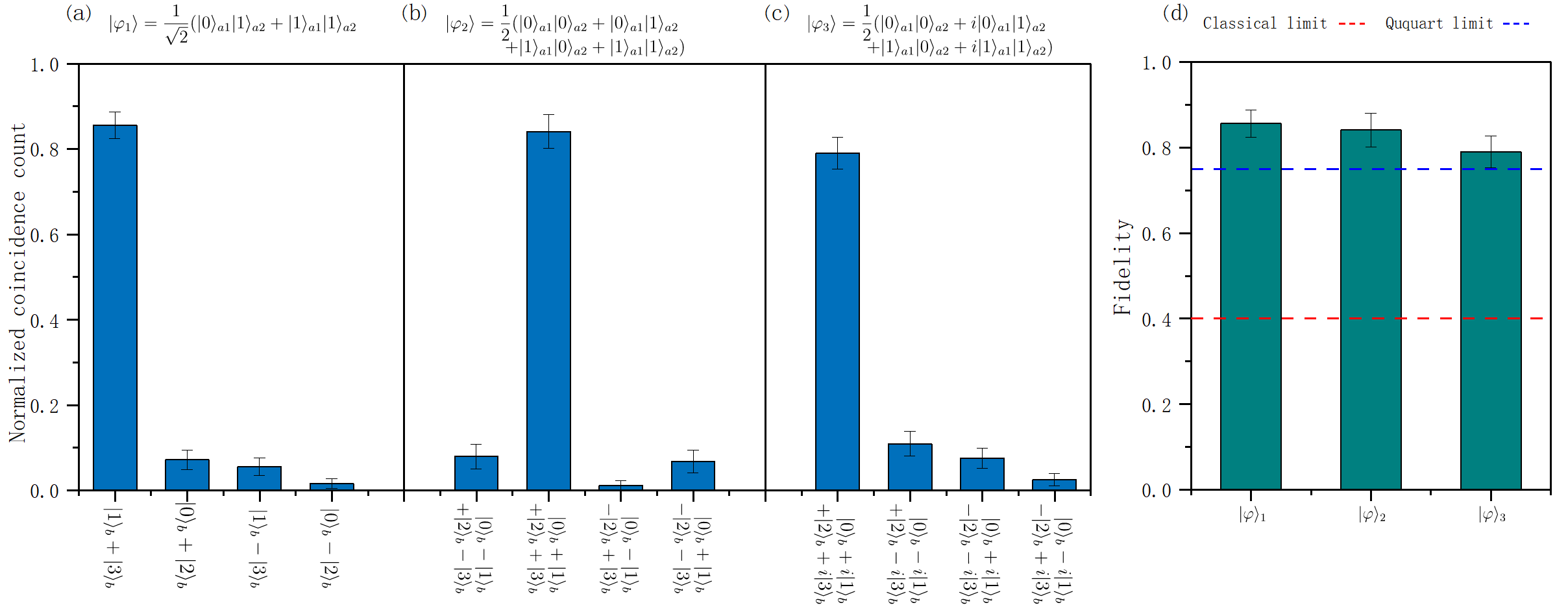}\\
	\captionsetup{justification=raggedright}
	\caption{Experimental results for the quantum teleportation from two qubits to a single ququart.
		(a-c) Measurement results of the final ququart state for the two-qubit states $|\varphi_1 \rangle$, $|\varphi_2 \rangle$, and $|\varphi_3 \rangle$. For photons a1 and a2, $|H\rangle$ and $|V\rangle$ are defined as $|0\rangle$ and $|1\rangle$, respectively. For photon b, $|Hu\rangle$,  $|Hl\rangle$, $|Vu\rangle$ and $|Vl\rangle$ are defined as $|0\rangle$, $|1\rangle$, $|2\rangle$ and $|3\rangle$, respectively. (d) Summary of the fidelities of the quantum teleportation for the three two-qubit states. The fidelities range from 0.79 to 0.86, which are well above both the optimal single-copy ququart state-estimation limit of $2/5$ and maximal qutrit-ququart overlap of $3/4$. The error bars (SD) are calculated according to propagated Poissonian counting statistics of the raw detection events.
		%
		%
	}
\end{figure*}

After preparing the (2, 2, 4) state, we use it as a resource to perform a proof-of-principle quantum teleportation experiment. 
As shown in Fig. 2, instead of sending photons $a1$ and $a2$ for qubit measurements (gray region), we send them into the teleportation setup (pink region). Photon $a1$ ($a2$) pass through a DoF converter consisting of a beam displacer BD3 (BD4) and the three half-wave plates (HWPs) before and after it, which expand the dimensionality of photon $a1$ ($a2$) from two to four dimensions, thus turning the three-photon state into
\begin{equation}
\begin{aligned}
	&\frac{1}{2}(|Hu\rangle _{a1} |Hu\rangle _{a2}|Hu\rangle_{b}+ |Hu\rangle _{a1}|Hl\rangle _{a2}|Hl\rangle_{b}\\
	&+|Hl\rangle _{a1} |Hu\rangle _{a2}|Vu\rangle_{b}+ |Hl\rangle _{a1}|Hl\rangle _{a2}|Vl\rangle_{b}).
\end{aligned}
\end{equation}
If the polarization and path DoFs of photons $a1$ and $a2$ are each written separately as a qubit, this three-photon state can be rewritten as
\begin{equation}
\begin{aligned}
	|H\rangle _{a1}|H\rangle _{a2}\otimes\frac{1}{2}(&|u\rangle _{a1} |u\rangle _{a2}|Hu\rangle_{b}+ |u\rangle _{a1}|l\rangle _{a2}|Hl\rangle_{b}\\
	+&|l\rangle _{a1} |u\rangle _{a2}|Vu\rangle_{b}+ |l\rangle _{a1}|l\rangle _{a2}|Vl\rangle_{b}),
\end{aligned}
\end{equation}	
where $u$ and $l$ denote photons in the upper and lower paths, respectively. Photon $a1$ ($a2$) then passes through HWP3 and quarter waveplate QWP3 (HWP4 and QWP4), and the two polarization qubits of photons $a1$ and $a2$ are then prepared  to
\begin{equation}
\alpha|H\rangle _{a1}|H\rangle _{a2}+\beta |H\rangle _{a1}|V\rangle _{a2}
	+\gamma |V\rangle _{a1}|H\rangle _{a2}+\delta |V\rangle _{a1}|V\rangle _{a2},
\end{equation}	
which is the two-qubit state that is set to be teleported.

The joint quantum state of photons $a1$, $a2$ and $b$ can now be written as
\begin{equation}
\begin{aligned}
	(\alpha&|H\rangle _{a1}|H\rangle _{a2}+\beta |H\rangle _{a1}|V\rangle _{a2}\\
	+\gamma &|V\rangle _{a1}|H\rangle _{a2}+\delta |V\rangle _{a1}|V\rangle _{a2})\\
	\otimes\frac{1}{2}(&|u\rangle _{a1} |u\rangle _{a2}|Hu\rangle_{b}+ |u\rangle _{a1}|l\rangle _{a2}|Hl\rangle_{b}\\
	+&|l\rangle _{a1} |u\rangle _{a2}|Vu\rangle_{b}+ |l\rangle _{a1}|l\rangle _{a2}|Vl\rangle_{b}).
\end{aligned}
\end{equation}	
It can be seen that there is a one-to-one correspondence between Eq. (10) and Eq. (4). Therefore, according to the previous theoretical derivation, implementing Bell state measurements of the polarization and path qubits of photon $a1$ and $a2$, respectively, and then sending the results to Bob for him to transform photon $b$ accordingly can realize the desired quantum teleportation.
%
Implementing a full deterministic quantum teleportation requires the use of active feed-forward, however, in this proof-of-principle experiment, we did not apply feed-forward but used post-selection to realize a probabilistic quantum teleportation.

We let photon $a1$ ($a2$) pass through BD5 (BD6), the subsequent waveplates and the polarization beamsplitter (PBS)  to be detected, thus realizing the projection of the polarization and path qubits of photon $a1$ ($a2$) onto the Bell state $\frac{1}{\sqrt{2}}\left( |H\rangle _{a1}|u\rangle _{a1}+|V\rangle _{a1}|l\rangle _{a1} \right)$ ($\frac{1}{\sqrt{2}}\left( |H\rangle _{a2}|u\rangle _{a2}+|V\rangle _{a2}|l\rangle _{a2} \right)$).
After completing the projection of photons $a1$ and $a2$, the quantum state of photon $b$ becomes
%
\begin{equation}
	\alpha |Hu\rangle_b +\beta |Hl\rangle_b +\gamma |Vu\rangle_b +\delta |Vl\rangle_b.
\end{equation}
Comparing Eq. (11) with Eq. (9), it can be seen that the two quantum states possess exactly the same form except for the difference in the state basis, so that the two-qubit state originally encoded in photons $a1$ and $a2$ has been transferred to photon $b$, thus realizing a quantum teleportation from two qubits to a ququart.

%

%

In our experiment, we have chosen three different two-qubit states for quantum teleportation, 
\[
\begin{aligned}
	|\varphi_1 \rangle=\frac{1}{\sqrt{2}}&(|H\rangle _{a1}|V\rangle _{a2}+|V\rangle _{a1}|V\rangle _{a2}, \\
	|\varphi_2 \rangle=\frac{1}{{2}}(&|H\rangle _{a1}|H\rangle _{a2}+|H\rangle _{a1}|V\rangle _{a2}\\
	+ &|V\rangle _{a1}|H\rangle _{a2}+ |V\rangle _{a1}|V\rangle _{a2}), \\
	|\varphi_3 \rangle=\frac{1}{{2}}(&|H\rangle _{a1}|H\rangle _{a2}+i|H\rangle _{a1}|V\rangle _{a2}\\
	+& |V\rangle _{a1}|H\rangle _{a2}+ i|V\rangle _{a1}|V\rangle _{a2}),
	\end{aligned}
\]
After completing the quantum teleportation of $|\varphi_1 \rangle$, $|\varphi_2 \rangle$ and $|\varphi_3 \rangle$, we measured photon $b$ at specific state bases to obtain the fidelities of these three quantum teleportations as $0.86\pm 0.03$, $0.84\pm 0.04$ and $0.79\pm 0.04$, respectively, as shown in Fig. 4(a)-(c).

The main sources of error in our experiment include double pair emission, imperfection in preparation of the initial states, and the non-ideal interference at the PPBS and BDs. 
Despite the experimental noise, the fidelities we obtained still far exceeds the classical limit of $2/5$, which is the best result that can be achieved using classical strategies \cite{Bru1999Optimal,hayashi2005reexamination}.
To demonstrate that the quantum teleportation of a four-dimensional quantum state is truly accomplished, the transferred quantum state needs to be uninterpretable by a mixture of superposition of three-dimensional or lower-dimensional quantum states, thus requiring that the average fidelity of the transferred quantum states with respect to the ideal states exceeds the ququart limit of $3/4$ \cite{luo2019quantum}.
The obtained average fidelity $0.83\pm 0.04$ also exceeds this bound, thus decisively proving that we have realized a four-dimensional quantum teleportation.

To summarize, we have experimentally prepared an asymmetric maximally entangled state consisting of two qubits and one ququart, and used it as a resource to implement a proof-of-principle quantum teleportation experiment, which realizes the transfer of an unknown four-dimensional quantum state from two qubits to a ququart. The asymmetric entangled state realized here has the potential to be widely used as a quantum interface in future quantum networks, enabling quantum information transfer between quantum objects of different dimensions through the quantum teleportation protocol demonstrated in this work.

\begin{acknowledgments}
	This work was supported by the National Key Research and Development Program
	(Grant No. 2017YFA0305200), the Key Research and Development Program of Guangdong Province of
	China (Grants No. 2018B030329001 and No. 2018B030325001), the National Natural Science Foundation
	of China (Grant No. 61974168). X.-Q. Zhou acknowledges support from the National Young 1000 Talents
	Plan. 
\end{acknowledgments}

\end{CJK}	
\end{document}